\documentclass[superscriptaddress,aps,prx,reprint,twocolumn,longbibliography,9pt]{revtex4-2}

\usepackage[english]{babel}
\usepackage{graphicx}
\usepackage{wrapfig}
\usepackage{color}
\usepackage{latexsym,amsmath}
\usepackage{physics}
\usepackage{dsfont}
\usepackage{chemformula}
\usepackage{tabularx}
\usepackage{float}
\usepackage{siunitx}
\usepackage{amssymb}
\usepackage{multirow}
\usepackage[permil]{overpic}

\usepackage{fancyhdr} 
\definecolor{cool}{RGB}{32,101,171}
\definecolor{hot}{RGB}{177,23,42}
\usepackage[pdftex,colorlinks=true, pdfstartview=FitV, linkcolor= hot, citecolor= hot, urlcolor= cool, hyperindex=true,hyperfigures=true]{hyperref}

\makeatletter
\renewcommand*{\fnum@figure}{{\normalfont\bfseries \figurename~\thefigure}}
\renewcommand*{\fnum@table}{{\normalfont\bfseries \tablename~\thetable}}
\makeatother

\usepackage{lipsum}
\usepackage{comment}
\definecolor{color_comment}{rgb}{0.8, 0.0, 0.0}
\definecolor{color_out}{rgb}{0.7, 0.7, 0.7}
\definecolor{color_suggestion}{rgb}{0.01, 0.75, 0.24}
\definecolor{color_todos}{rgb}{0.0, 0.5, 1.0}

\newif\ifdraft

\drafttrue 

\newcommand{\ustuttgartIII}{Institute for Theoretical Physics III and Center for Integrated Quantum Science and Technology,
University of Stuttgart, 70550 Stuttgart, Germany}
\newcommand{\ustuttgartV}{5th Institute for Physics and Center for Integrated Quantum Science and Technology,
University of Stuttgart, 70550 Stuttgart, Germany}
\newcommand{\uulm}{Institute for Complex Quantum Systems, University of Ulm, D-89081 Ulm, Germany}
\newcommand{\unipd}{Dipartimento di Fisica e Astronomia “G. Galilei” \& Padua Quantum Technologies Research Center, Università degli Studi di Padova, I-35131 Padova, Italy}
\newcommand{\pdinfn}{INFN Istituto Nazionale di Fisica Nucleare, Sezione di Padova, I-35131 Padova, Italy}

\makeatletter
\newcommand\thefontsize{The current font size is: \f@size pt}
\makeatother

\begin{document}

\title{Error-budgeting for a controlled-phase gate with strontium-88 Rydberg atoms}

\author{Alice Pagano}
\affiliation{\uulm}

\author{Sebastian Weber}
\affiliation{\ustuttgartIII}

\author{Daniel Jaschke}
\affiliation{\uulm}
\affiliation{\pdinfn}

\author{Tilman Pfau}
\affiliation{\ustuttgartV}

\author{Florian Meinert}
\affiliation{\ustuttgartV}

\author{Simone Montangero}
\affiliation{\uulm}
\affiliation{\pdinfn}
\affiliation{\unipd}

\author{Hans Peter B{\"u}chler}
\affiliation{\ustuttgartIII}

\date{\today}

\begin{abstract}
We study the implementation of a high fidelity controlled-phase gate in a Rydberg quantum computer. The protocol is based on a symmetric gate with respect to the two qubits as  experimentally realized by Levine {\it et al} [Phys. Rev. Lett. 123, 170503 (2019)], but allows for arbitrary pulse shapes with time-dependent detuning.  Optimizing the pulse shapes, we introduce laser pulses which shorten the time spent in the Rydberg state by 10\% and 
reduce the leading contribution to the gate infidelity, i.e., the decay from the Rydberg state.
Remarkably, this reduction can be achieved for smooth pulses in detuning and smooth turning on of 
the Rabi frequency as required in any experimental realization. 
We carefully analyze the influence of fundamental error sources such as the photon recoil, the microscopic interaction potential, as well as the harmonic trapping of the atoms for an experimentally realistic setup based on strontium-88 atoms. We find that an average gate fidelity above 99.9\% is possible for a very conservative estimation of experimental parameters.
\end{abstract}

\maketitle

\section{Introduction}
\label{sec:introduction}

Arrays of identical neutral atoms trapped in optical tweezers enable several attractive features for the realization of universal quantum computers. 
Especially, qubits encoded into two different internal states of the atom are inherently identical and exhibit long coherence times. In addition, the ability to scale the number of trapped atoms up to several hundred in such arrays has been demonstrated~\cite{scholl2021quantum,ebadi2021quantum}. 
A promising approach for the implementation of two-qubit and multi-qubit gates is to temporarily excite the atoms into Rydberg states, while  accurate local control and single-qubit gates are realized by microwave or optical transitions~\cite{saffman2010,Saffman2016,henriet2020,wu2021}. Several remarkable breakthroughs towards the implementation of the Rydberg quantum computing platforms have been achieved~\cite{graham2022,bluvstein2021quantum}, but still, the fidelities for the two-qubit gates are lower than for competitive platforms such as ion traps~\cite{Gaebler2016,Ballance2016}. 
In this manuscript, we perform a careful analysis for two-qubit controlled-phase gates, and demonstrate an 
approach towards the realization of high fidelity gates with strontium atoms.

The principal idea for the realization of two-qubit controlled-phase gates between neutral 
atoms~\cite{jaksch2000} is based on the so-called \emph{Rydberg blockade}, where the strong interaction
between two Rydberg states quenches the simultaneous excitation of nearby Rydberg atoms on distances up 
to several micrometers. Motivated by the first proof-of-principle experimental demonstrations 
\cite{Urban2009,Gaetan2009}, a large theoretical interest was focused on improving the gate fidelity and finding
alternative protocols~\cite{Muller2014,Han2016,Su2016,Beterov2016,Petrosyan2017,Saffman2020,Mitra2020,Fu2021}, 
as well as extending these ideas to multi-qubit gates~\cite{Muller2009,Isenhower2011,Beterov2018,Khazali2020}.
However, very few studies have included the effect of photon recoil and the harmonic trapping 
of the atoms~\cite{saffmann2021}. On the experimental side, the tremendous progress in manipulating 
individual Rydberg atoms in optical tweezers~\cite{browaeys2020} has led to the implementation 
of a controlled-phase gate with a Bell-state fidelity greater than 97.4\% in rubidium
atoms~\cite{levine2019}. Furthermore, these ideas have been extended to alkaline-earth atomic
species such as  strontium~\cite{stellmer2014, cooper2018, norcia2018, kanungo2020, teixeira2020, qiao2021}, and  the generation of Bell states with a fidelity greater than 99.1\% has been demonstrated~\cite{madjarov2020}. These exciting results show how Rydberg gates can achieve very high fidelities, exploiting
fast protocols with a typical gate time of the order of a sub-microsecond. Recently,
the execution of quantum algorithms has been demonstrated on a programmable neutral atom processor highlighting the emergent capability of these devices for programmable 
quantum computation~\cite{graham2022} as well as the ability to change the connectivity between the 
qubits during the execution of the quantum algorithm~\cite{bluvstein2021quantum}.

In this paper, we optimize the laser pulse shapes for the realization of a symmetric controlled-phase 
gate between neutral atoms, and perform  an analysis of the different fundamental error sources.  The 
method is motivated by the gate protocol presented in Ref.~\cite{levine2019},
but allows for an arbitrary time-dependent detuning instead of a sharp phase jump.  
In a realistic experimental setup, a finite bandwidth and smoothness of the pulse shapes is 
important in order to suppress excitations into additional Rydberg states as well as experimental limitations to control laser pulses.
In addition, the leading mechanism limiting the gate fidelity is the finite lifetime of the Rydberg 
states, and shortening the time spent in the Rydberg state for a fixed maximal Rabi
frequency is crucial. We find that several different pulse shapes  provide a $10\%$ improvement compared to the 
original proposal of Ref.~\cite{levine2019}. We also confirm the optimality of those pulse shapes employing quantum optimal control optimization~\cite{Koch2016, Glaser2015,muller2021}.
Among those shapes are also very smooth ones, i.e., with limited bandwidth, and thus depending on a very limited number of parameters~\cite{Lloyd2014a}. In the 
next step, we perform a careful analysis of the different contributions limiting the gate fidelity such 
as the finite lifetime of the Rydberg state, the photon recoil of the excitation laser,  the microscopic 
interaction potential between the Rydberg states, as well as the harmonic trapping in three-dimensions of 
the atoms, and imperfect cooling in the motional ground state. For an experimentally realistic setup with
strontium-88 atoms, we demonstrate that an average gate fidelity above $99.9\%$ is achievable, and analyze 
the contribution of each of the above phenomena to the infidelity. We find that 
the effect of the photon recoil energy is reduced for weak harmonic trapping frequencies, and 
provide a simple analytical explanation to understand this behavior.

\section{Controlled-phase gate implementation}
\label{sec:different_shapes}

\begin{figure}[t]
    \centering
    \includegraphics[width=0.45\textwidth]{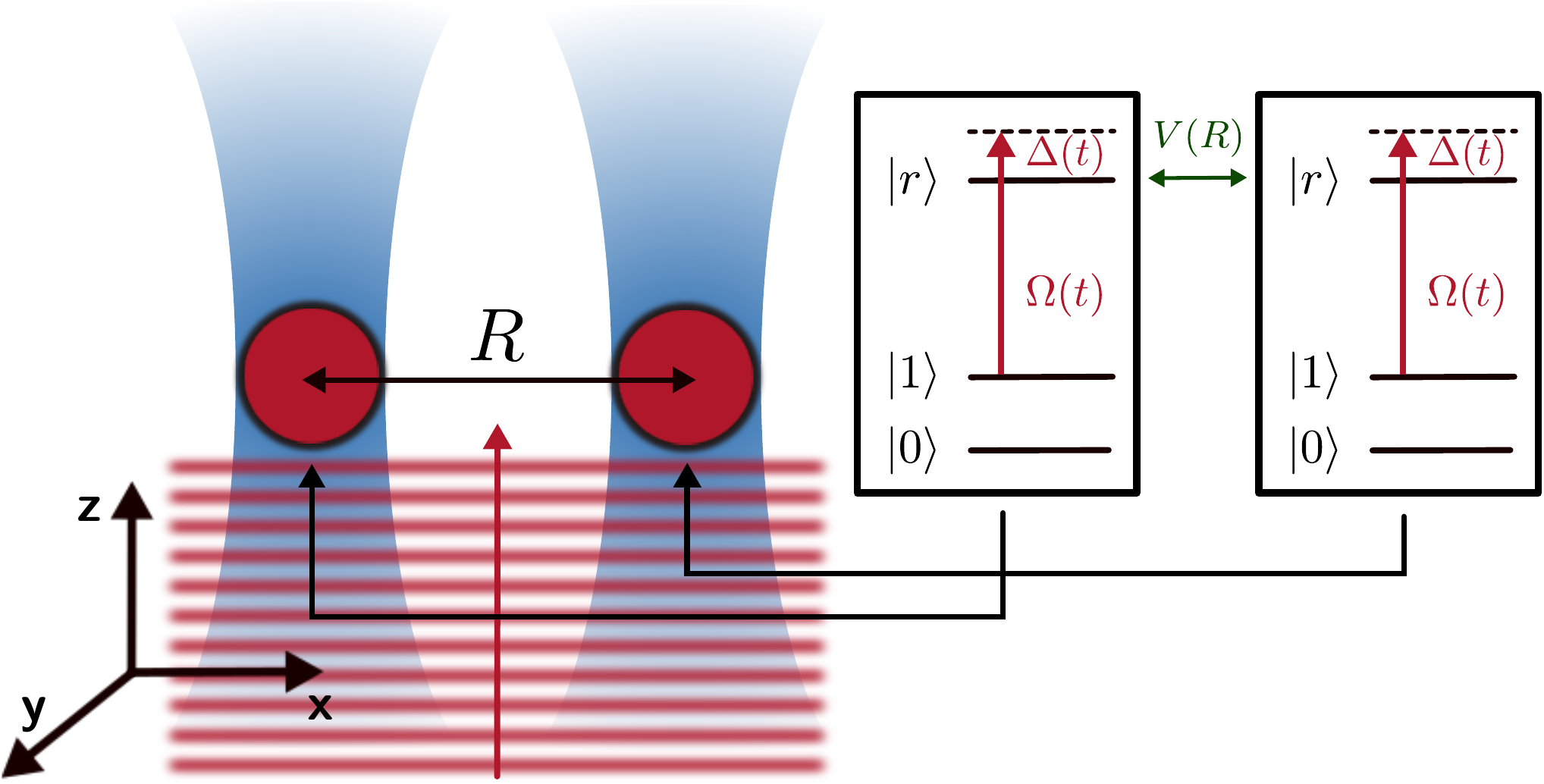}
    \caption{\emph{Setup for the realization of a controlled-phase gate with neutral atoms.} The two atoms are trapped in optical tweezers (in blue) at a fixed distance $R$ along the $x$-direction. In the idealized description, we consider a three-level structure with states $\ket{0}$, $\ket{1}$ and the Rydberg state $\ket{r}$ for each atom. The Rydberg state is coupled with the state $\ket{1}$ by homogeneously driving a global laser, i.e., a plane wave (in red), with Rabi frequency $\Omega(t)$ and detuning $\Delta(t)$ along the $z$-direction. The Rydberg states interact via a van der Waals interaction $V(R)$ (green arrow) which strongly depends on the distance between the atoms.}
    \label{fig:setup}
\end{figure}

In this section, we provide a detailed description of the controlled-phase gate protocols. The qubit 
states $\ket{0}$ and $\ket{1}$ are encoded into two internal states of the neutral atom, and each atom 
is individually trapped in three dimensions by an optical tweezer. For the two-qubit gate, the configuration
is illustrated in  Fig.~\ref{fig:setup}: the two neutral atoms are  at a fixed distance $R$ along the
$x$-dimension. The gate is achieved by coupling the logical state $\ket{1}_{i}$ of the $i^{\mathrm{th}}$
qubit to a strongly interacting Rydberg state $\ket{r}_{i}$ with time-dependent Rabi frequency $\Omega(t)$
and time-dependent detuning $\Delta(t)$; here, we are interested in the implementation of a symmetric gate,
where the coupling to the Rydberg state as well as the detuning acts on both atoms equally~\cite{levine2019}.
The  laser pulse is characterized by the overall duration $\tau$ and the maximal Rabi frequency $\Omega_{0}$. The strong interaction between 
the Rydberg level of each qubit then allows for the realization of a controlled-phase gate. The idealized description of the 
gate protocol is therefore achieved by including for each neutral atom the three states $\ket{0}_{i}$, $\ket{1}_{i}$, and $\ket{r}_{i}$; the Hamiltonian governing the dynamics takes the form
\begin{equation}                                                                        \label{eq:simple_hamiltonian}
  H= H_{0} + H_{\rm \scriptscriptstyle int} \, .
\end{equation}
The term $H_{0}$ accounts for the coupling to the Rydberg state by the driving laser pulse 
within the rotating frame and applying the rotating wave approximation,
\begin{equation}
    H_0= \hbar \sum_{i=1}^{2}   \left[ \frac{\Omega(t)}{2} \Big( \sigma_{i}^{+}+ \sigma_{i}^{-} \Big) - \Delta(t) n_{i} \right] \, ,
    \label{eqn:h0simple}
\end{equation}
with  $\sigma_{i}^{+}=|r\rangle \langle 1|_{i}$,  $\sigma_{i}^{-}=|1\rangle \langle r|_{i}$  
and $n_{i}= |r\rangle\langle r |_{i} $.
The interaction between the Rydberg states reduces to
\begin{equation}                                                                      \label{eq:hintsimple}
    H_{\rm \scriptscriptstyle int}= V n_{1} n_{2}.
\end{equation}
Herein, $V$ denotes the interaction strength between the two Rydberg states, which in our setup  
is determined by van der Waals interactions. 
We discuss multiple corrections to this idealized description for a realistic experimental realization in Sec.~\ref{sec:error-budgeting}.

The idea for the realization of a two-qubit phase gate is to work in the regime of a strong Rydberg blockade, i.e., $V/ \hbar \Omega_{0} \gg 1$ ~\cite{browaeys2020}. We analyze the gate's behavior via the driving Hamiltonian $H$ on the four computational basis states. Since the state $\ket{00}$
is uncoupled by the Rydberg laser, it does not evolve and trivially maps the initial state $\ket{00}$ onto the final state $\ket{00}$.  If one of the two atoms is in the logical state $\ket{0}$, only the qubit in the logical state $\ket{1}$ undergoes a non-trivial time evolution. The dynamics can be described with a two-level system with states $\ket{1}$ and $\ket{r}$, e.g., with the two-qubit states $\ket{10}$ and $\ket{r0}$ when the second qubit is in the $\ket{0}$ state.
Note that the states $\ket{10}$ and $\ket{01}$ exhibit equivalent dynamics in this symmetric setup. Instead, the dynamics of the initial state $\ket{11}$ follows a two-level system with states $\ket{11}$ and $(\ket{1r}+\ket{r1}) / \sqrt{2}$ thanks to the strong Rydberg blockade and displays a collectively enhanced Rabi frequency $\sqrt{2}\Omega(t)$.
The first requirement on the laser pulses is therefore to guarantee that both states end up again in the logical qubit states, i.e., avoid leakage to the states $\ket{r}$ outside of the computational space. The phases acquired during the laser pulse depend on the dynamical phase as well as the Berry phase  acquired. However, the enhancement in Rabi frequency  leads to different trajectories on the Bloch sphere, and therefore, the state $\ket{11}$ picks up a phase $\phi_{11}$, which is different from the phase $\phi_{10}\equiv\phi_{01}$ picked up by the two states $\ket{10}$ and $\ket{01}$. Therefore, the second requirement on the laser pulses is to fix the phases to satisfy the condition
\begin{equation}
    \phi_{11}-\phi_{01}-\phi_{10} = (2n+1)\pi,
    \label{eq:gate-condition}
\end{equation} 
with $n$ an integer. Then, the implemented two-qubit gate corresponds to a controlled-phase between the two qubits 
up to single qubit rotations $\ket{0}_{i} \rightarrow \ket{0}_{i}$
and $\ket{1}_{i} \rightarrow e^{\mathrm{i} \phi_{10}} \ket{1}_{i}$.  These conditions are also achieved for realistic imperfect blockade with a finite interaction strength $V$ between the two
Rydberg states; we discover a huge variety of intuitive laser pulses for the realization of the phase gate, see Fig.~\ref{fig:benchmark_shapes}(a) for a selection of shapes.

The characteristic time scale for the phase gate is determined by the maximal Rabi frequency $\Omega_{0}$ of the driving lasers; in current experimental setups, the latter is restricted to 
the range of approximately $2 \pi \cdot 10 {\rm MHz}$. In combination with this restriction, the finite lifetime $1/\gamma$ of the Rydberg states provides a fundamental limitation on the fidelity achievable in a Rydberg quantum gate. Therefore, the leading step to improve the gate fidelity is to find the  pulse sequence  in $\Omega(t)$ and $\Delta(t)$  with a fixed maximal Rabi frequency $\Omega_{0}$, where the time spent in the Rydberg level is minimized. We define the latter as the time-integrated probability to be in Rydberg state $T_{r}^{\alpha}$ and consider moreover the mean value $\overline{T}_{r}$
\begin{align}
    T_{r}^{\alpha} &= \int_{0}^{\tau} \text{d}t \sum_{i=1}^{2} \langle n_{i}(t) \rangle \, ,                       \label{eq:time_rydberg_state} \\
    \overline{T}_{r} &= \frac{1}{3} \left( T_{r}^{\ket{01}} + T_{r}^{\ket{10}} + T_{r}^{\ket{11}} \right) \, ,             \label{eq:mean_time_rydberg_state}
\end{align}
where we recall that the operator $n_{i}$ measures the population in the Rydberg state.
The superscript $\alpha$ identifies the initial state before applying the gate, i.e., $\alpha$ equal to $\ket{11},\ket{01}$ or $\ket{10}$. The integration covers the total time for the Rydberg gate $\tau$. 
Furthermore, the restriction of the analysis to smooth pulse shapes is important, as a high bandwidth of the pulses is extremely challenging to achieve in the experimental setup, and also leads to excitations into other Rydberg levels not included in the idealized 3-level description.

\newcounter{subfig}\setcounter{subfig}{0}
\renewcommand\thesubfig{(\alph{subfig})}
\begin{figure*}[ht]
        \centering
        \begin{overpic}[width=\linewidth]{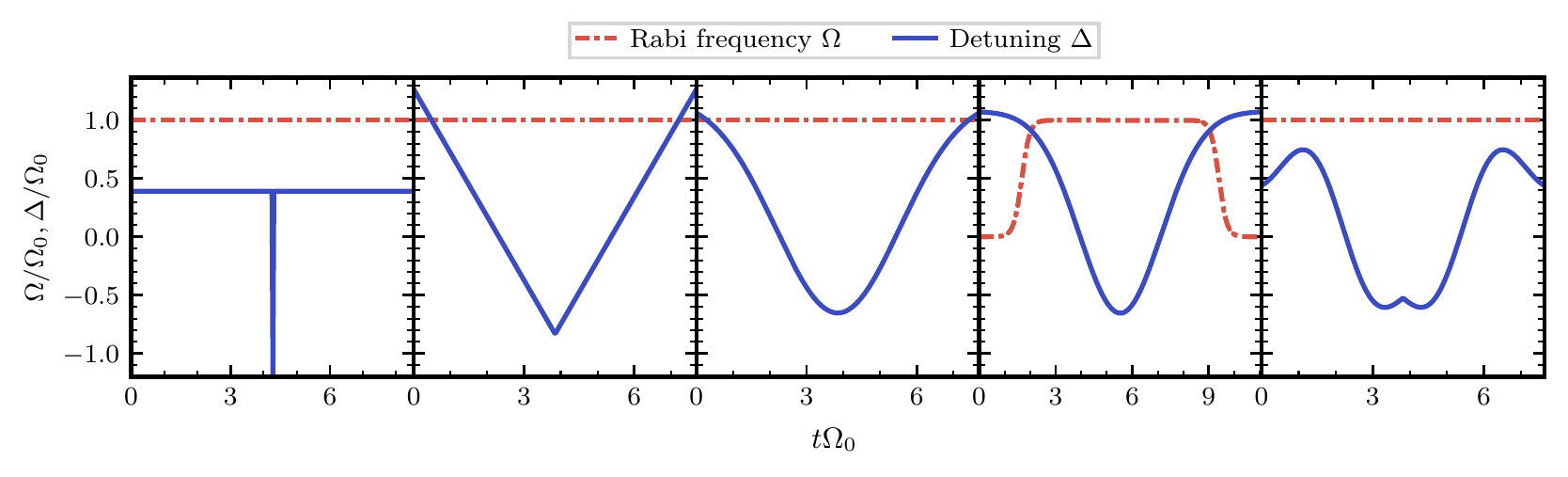}
            \put(15,285){\color{black}\stepcounter{subfig}\thesubfig}
            \put(100,90){(I)}
            \put(290,90){(II)}
            \put(460,90){(III.A)}
            \put(640,90){(III.B)}
            \put(820,90){(IV)}
        \end{overpic}%
        \vspace{8pt}
        \begin{overpic}[scale=0.64]{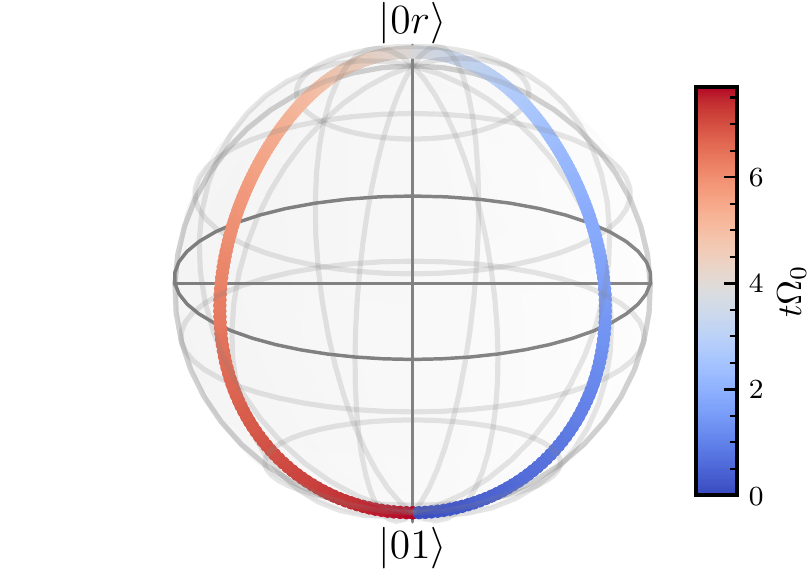}
            \put(15,700){\color{black}\stepcounter{subfig}\thesubfig}
        \end{overpic}%
        \hspace{1cm}
        \begin{overpic}[scale=0.64]{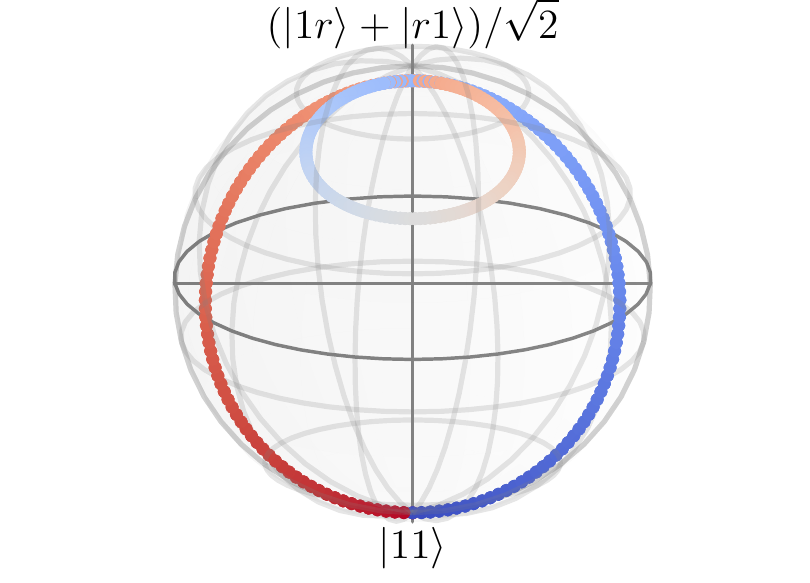}
            \put(15,700){\color{black}\stepcounter{subfig}\thesubfig}
        \end{overpic}%
        \hspace{-1.7cm}
        \begin{overpic}[scale=0.64]{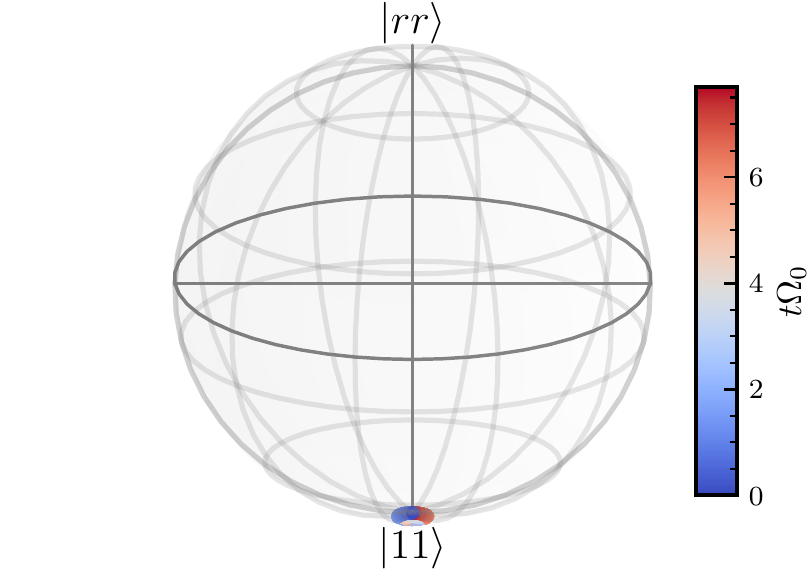}
        \end{overpic}%
        \vspace{8pt}
        \begin{overpic}[width=\linewidth]{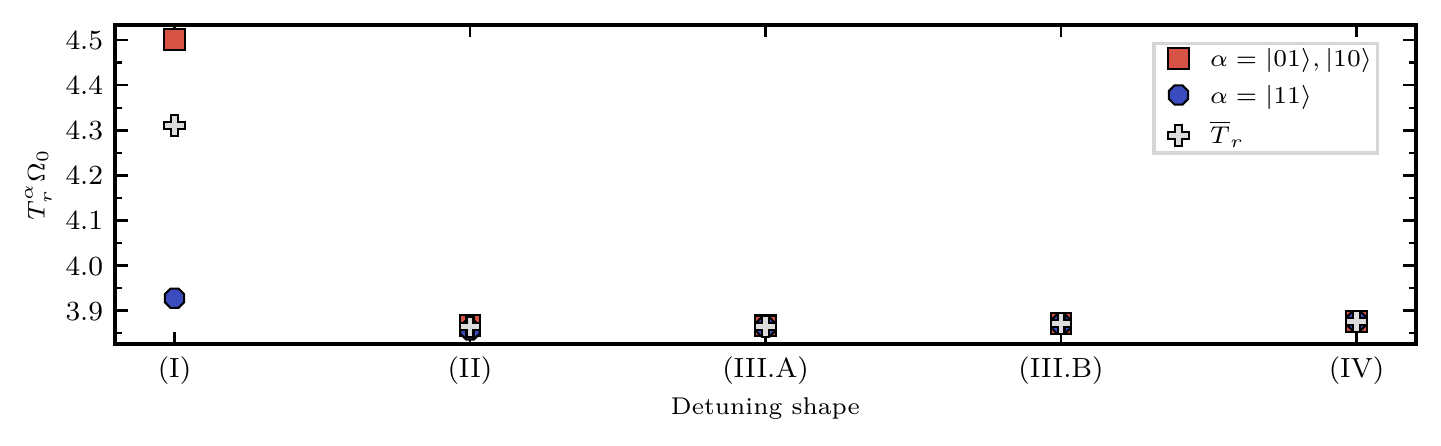}
            \put(15,305){\color{black}\stepcounter{subfig}\thesubfig}
            \put(150,100){\includegraphics[scale=0.7]{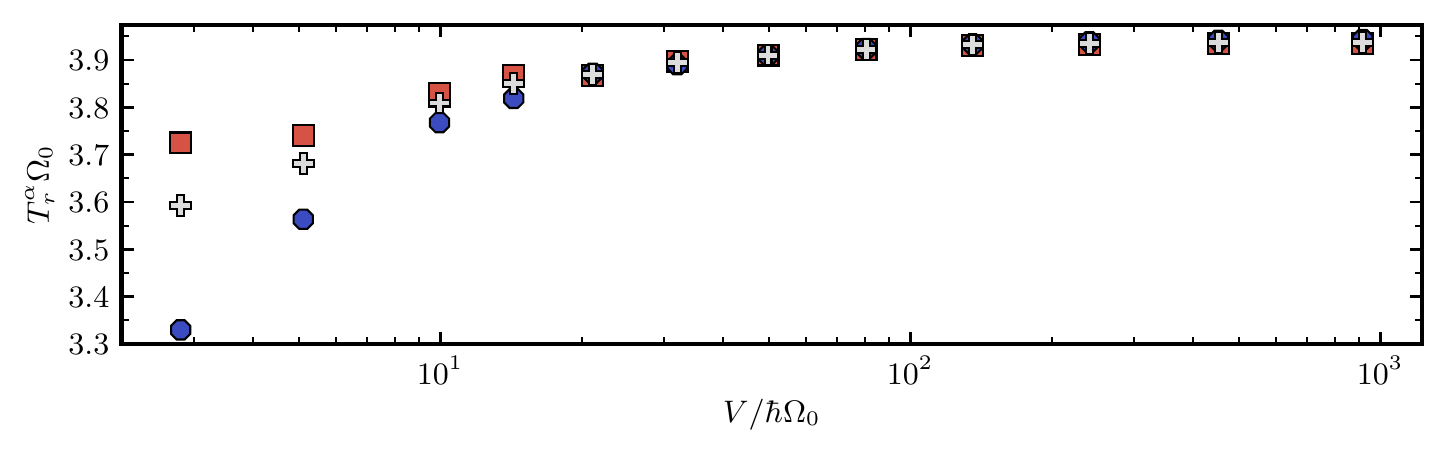}}
        \end{overpic}%
    \caption{\setcounter{subfig}{0}\emph{Analysis for the realization of a controlled-phase gate for two qubits via the interaction of their Rydberg states.} \protect\stepcounter{subfig}\thesubfig{}~The controlled-phase gate can be realized by using different time-dependent detuning shapes $\Delta(t)$ at interaction strength $V/\Omega_{0}\hbar = 21.1$. The shape (I) is a $\delta$-function as recently applied by Levine {\it et al} [Phys. Rev. Lett. 123, 170503 (2019)].
    The shape (II) is an isosceles triangle. The shape (III.A) is a Gaussian with width $w \Omega_{0}=1.7$. A solution with a Gaussian detuning can be found also introducing a finite rise time $\kappa \Omega_{0}=0.31$ on the Rabi frequency $\Omega(t)$ as one can see in  (III.B).
    Then, there is the solution (IV) obtained by using the optimal control algorithm dCRAB. 
    \protect\stepcounter{subfig}\thesubfig{}, \protect\stepcounter{subfig}\thesubfig{} Dynamics on the Bloch sphere for a controlled-phase gate realized by assuming the shape (III.A).
    \protect\stepcounter{subfig}\thesubfig{} Comparing the total time spent in the Rydberg state $T_{r}^\alpha$ for the different shapes by considering separately $\alpha=\ket{11}$ and $\alpha=\ket{01}$ as initial state for the evolution. The average time in the Rydberg state is $\overline{T}_r=( T_{r}^{\ket{01}} + T_{r}^{\ket{10}} + T_{r}^{\ket{11}})/3$. In order to show that we can drop the assumption of a perfect blockade, we also illustrate the total time spent in the Rydberg state as a function of the Rydberg interaction. 
    }
    \label{fig:benchmark_shapes}
\end{figure*}
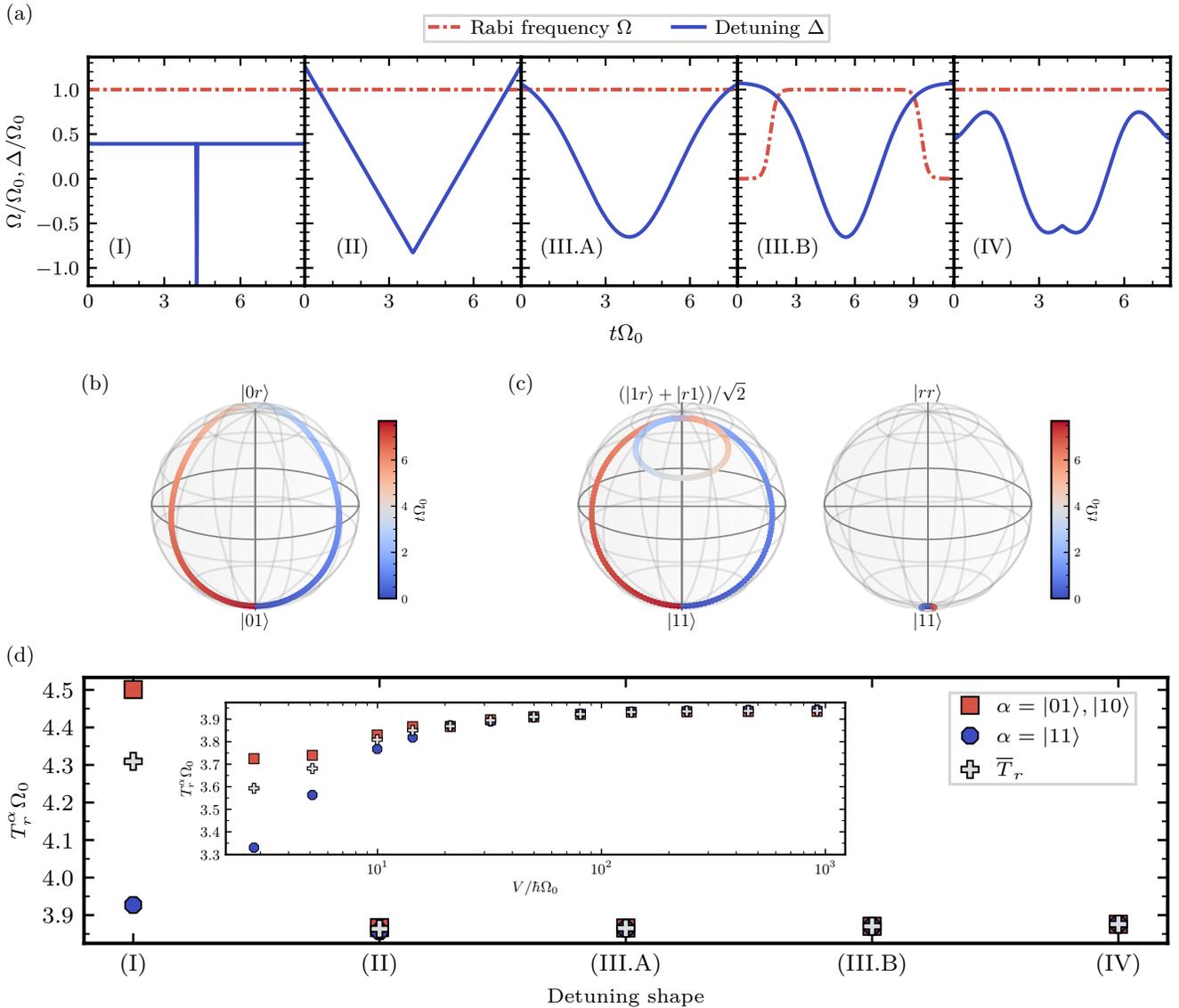

In the following, we show that the controlled-phase gate can be realized for a variety of different detuning shapes $\Delta(t)$ and Rabi frequencies $\Omega(t)$, and we optimize the pulse shapes via the Bell state fidelity towards a minimal time spent in the Rydberg state. For this analysis, we simulate the dynamics of the three-level Hamiltonian $H$ defined in Eqs.~\eqref{eq:simple_hamiltonian} to \eqref{eq:hintsimple} with the open-source software QuTiP~\cite{qutip2012}. We start with the original gate as proposed in Ref.~\cite{levine2019} with a constant Rabi frequency turned on at $t=0$ and turned off at $t=\tau$, see shape (I) in Fig.~\ref{fig:benchmark_shapes}(a). Note that the transformation $\ket{r} \rightarrow e^{\mathrm{i} \theta(t)} \ket{r}$ connects our approach with a time-dependent detuning to a setup with a time-dependent laser phase $\theta(t)$ via  
\begin{equation}
 \Delta(t)= \Delta_0 + \partial_{t} \theta(t)\;. 
\end{equation}
Then, the sharp phase jump of $\theta(t)$ at $t=\tau/2$ in Ref.~\cite{levine2019} corresponds to a  $\delta$-peak in the detuning; the infinite value of the $\delta$-function is evidently not fully visible in Fig.~\ref{fig:benchmark_shapes}(a) for shape (I).
For a van der Waals interaction strength $V/ \hbar \Omega_{0} = 21.1$, that can be achieved with realistic experimental parameters as described in Sec.~\ref{sec:error-budgeting}, the time for the gate becomes $\tau  \Omega_{0}= 8.53$ with a significant time spent in the Rydberg state. Figure~\ref{fig:benchmark_shapes}(d) indicates an average time in $\ket{r}$ of $\overline{T}_r \Omega_0 = 4.31$; on average, each atom is in the Rydberg state for approximately one-fourth of the gate duration $\tau\Omega_0$, where the maximum of $T_{r}^{\alpha}$ from Eq.~\eqref{eq:time_rydberg_state} is $2 \tau$, or $\tau$ in case of a perfect Rydberg blockade. 
This fraction of the time is significant and shows us why the finite lifetime of the Rydberg state is a substantial source of error for a Rydberg quantum gate and why its minimization is important.
Note that under realistic experimental conditions, the realization of a perfect $\delta$-peak is impossible, but requires the finite bandwidth to be taken into account, which leads to an additional intrinsic error. After the analysis of many optimal control solutions, we discover that the pulses are not overly complicated. Therefore, we turn to realizing the controlled-phase gate for pulse shapes with a few parameters to be optimized, e.g., height and width; first, we demonstrate the gate with an isosceles triangle detuning as in shape (II) from Fig.~\ref{fig:benchmark_shapes}(a).
In this case, the optimal gate takes $\tau \Omega_{0} = 7.69$, and is approximately $10\%$ faster
than the previous protocol (I), while reducing the time spent in the Rydberg states $ \overline{T}_{r}\Omega_0= 3.86$  
by approximately $10\%$, see Fig.~\ref{fig:benchmark_shapes}(d). Note that 
for the optimization we vary the shape's parameters for each pulse shape in order to minimize the gate 
duration with a gate infidelity less than $10^{-6}$. This gate time and time spent in the Rydberg state is very
robust to different pulse shapes. As an additional example, we study the Gaussian shape (III.A); the optimal gate duration is $\tau \Omega_{0} = 7.69$ with $ \overline{T}_{r} \Omega_0=3.86$ and a width  $w$ of the Gaussian 
shape $w \Omega_{0}=1.7$. Thus, the duration of the Gaussian and triangular protocols are
almost the same. Nevertheless, the major advantage of the Gaussian protocol lies in the better 
feasibility of the shape to be realized from an experimental point of view. Furthermore, we also 
study a realistic turning on of the Rabi frequency and Gaussian detuning with the shapes (III.B): we replace the sharp step function by a smooth pulse proportional to $\tanh(t/\kappa)$  
with the characteristic time scale $\kappa \Omega_{0}=0.31$. Then,
the gate naturally takes longer, but importantly the time spent in the Rydberg state remains 
essentially invariant under a smooth pulse shape of the Rabi frequency with $ \overline{T}_{r}
\Omega_0=3.87$. Finally, we present one optimal control analysis with an arbitrary shape for the 
detuning with the goal to further minimize the time spent in the Rydberg state.  We optimize the 
detuning pulse by using the optimal control algorithm dCRAB~\cite{muller2021,caneva2011,Rossignolo2021_soft}. Its key 
feature is to expand the detuning into a sum of truncated and randomized basis functions where we limit the frequency range; then, the 
problem is recast to a multi-variable function minimization that can be performed via direct-search
methods. 
The figure of merit chosen for the 
optimization is the Bell state fidelity, see Sec.~\ref{sec:error-budgeting}. Shape (IV) in
Fig.~\ref{fig:benchmark_shapes}(a) illustrates one of the optimal control solutions;  
we assume symmetric detuning around half the gate duration for the optimization. This symmetric
assumption produces the kink in the middle for the solution (IV). We enforce this constraint since our
initial optimal control pulses converge to a symmetric solution even without the symmetric assumption.
Thus, the constraint conveniently reduces 
the number of parameters to be optimized without affecting the success of the optimization. 
However, we find out that a further reduction of the gate time 
is impossible up to the given precision with respect to the Gaussian, 
or the triangular, protocol. Thus, these two protocols are already among the fastest possible solutions.
A minimal gate time is in agreement with the quantum speed limit for the transition between
states, although the actual value is more subtle to evaluate~\cite{caneva2009,MooreTibbetts2012}. There are protocols that avoid passing through the state $\ket{0r}$ in contrast to Fig.~\ref{fig:benchmark_shapes}(b), and therefore, the
Rabi oscillation between $\ket{01}$ and $\ket{0r}$ cannot be directly used to estimate the quantum speed limit.

 \begin{figure}[t]
    \centering
    \includegraphics{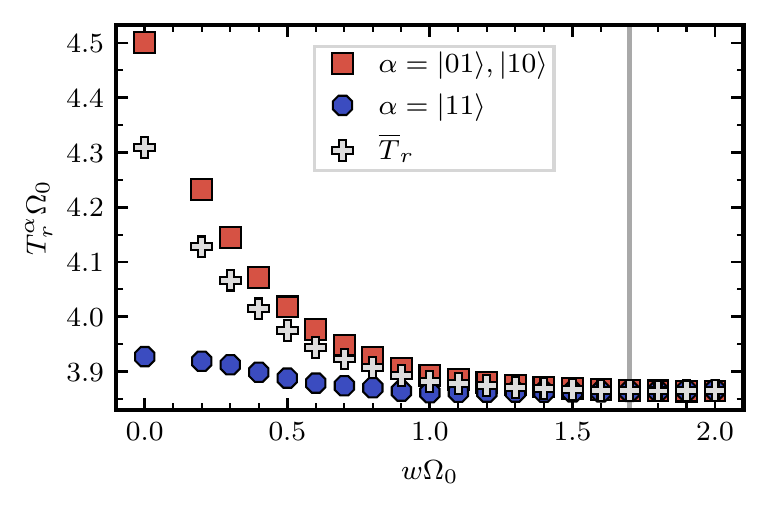}
    \caption{\emph{The time spent in the Rydberg state $T_{r}^{\alpha}$ as a function of the Gaussian width $w \Omega_{0}$ by assuming a Gaussian detuning protocol}. Each data point optimizes the peak's center and height after fixing the width of the Gaussian and is the solution with minimal infidelity, which is zero up to numerical precision. The point at $w=0$ is obtained with the $\delta$-function protocol. The grey line is the Gaussian width $w\Omega_0 = 1.7$ of the best pulse (III.A).
    }
    \label{fig:lukin_connection}
\end{figure}

The dynamics on the Bloch sphere for the Gaussian pulse shape (III.A) are shown in
Fig.~\ref{fig:benchmark_shapes}(b) and (c). As expected, the states $\ket{01}$, $\ket{10}$, and $\ket{11}$ 
return to themselves at the end of the evolution. Figure~\ref{fig:benchmark_shapes}(b) concentrates on the 
two-level dynamics when the first qubit is in the state $\ket{0}$; the dynamics for the initial state 
$\ket{10}$ follows the same evolution, respectively. The dynamics from the initial state $\ket{11}$ is 
highlighted in two cases in Fig.~\ref{fig:benchmark_shapes}(c), where the Bloch sphere for the states 
$\ket{11}$ and $(\ket{1r} + \ket{r1}) / \sqrt{2}$ covers the main part of the dynamics, while the right part 
describes the effects from an imperfect Rydberg blockade. In contrast to the original protocol in 
Ref.~\cite{levine2019} with a phase jump, the state $\ket{11}$ does not return to $\ket{11}$ at $t=\tau/2$. 
Note that the imperfect blockade with $V/ \hbar \Omega_{0} = 21.1$ also leads to a finite probability of 
both atoms being excited to the Rydberg state, i.e., state $\ket{rr}$, for the dynamics of the initial state 
$\ket{11}$. We define the integrated time in the state $\ket{rr}$ analog to 
Eq.~\eqref{eq:time_rydberg_state} as $T_{rr} = \int \mathrm{d}t \langle n_{1} n_{2} \rangle$, which is around
$ T_{rr} \Omega_0 = 0.0045$ in this case. Note that we can estimate this time within second order perturbation theory for strong interactions as $T_{rr} \approx (\hbar \Omega_{0}/\sqrt{2}V)^2 \: T_{r}^{\alpha}$ with $\alpha=\ket{11}$. 
For strong interactions $V/\hbar \Omega_{0} > 10$, the Rydberg times very quickly tend towards the value of perfect 
blockade, see inset of Fig.~\ref{fig:benchmark_shapes}(d), while for weaker interactions $V/\hbar \Omega_{0} < 10$ 
we can not always find a solution for given threshold on the infidelity $< 10^{-6}$, but surprisingly the time spent in the Rydberg state decreases for the interaction strengths with a solution.

The Gaussian protocol allows us to smoothly connect the optimal pulse to the original
implementation with a sharp phase jump~\cite{levine2019}: a phase jump corresponds to a $\delta$-function 
in the detuning and can be realized by decreasing width $w$ of the Gaussian shape (III.A).  
In Fig.~\ref{fig:lukin_connection}, we show the total time spent in the Rydberg state $T_{r}^{\alpha}$ and $\overline{T}_{r}$ as a function of the width $w$ of the Gaussian. 
One can observe from the plot that the time spent in the Rydberg state approaches  the one of the $\delta$-function protocol in the limit of small Gaussian width. 

\begin{figure}[t]
    \centering
    \begin{overpic}{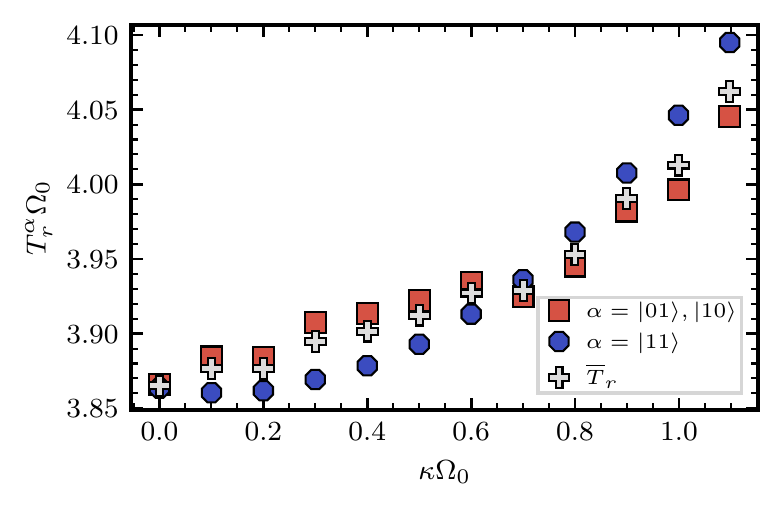}
        \put(650,450){\color{black}\vector(2,1){250}}
        \put(170,265){\includegraphics[scale=0.8]{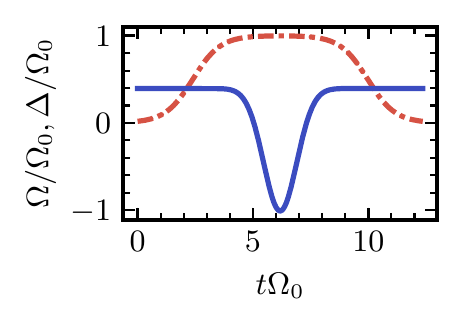}}
    \end{overpic}%
    \caption{\emph{The time spent in the Rydberg state $T_{r}^{\alpha}$ as a function of the characteristic time scale $\kappa \Omega_{0}$ by assuming a Gaussian detuning protocol with a realistic turning on of the Rabi frequency}. The inset shows the pulse shape for the detuning (blue) and the Rabi frequency (red) for the slowest turning on with $\kappa \Omega_{0}=1.2$.
    }
    \label{fig:raise_time_analysis}
\end{figure}

Finally, we evaluate the time spent in the Rydberg state for increasing characteristic time scale $\kappa$ to turn on the Rabi frequency for the pulses like (III.B), see Fig.~\ref{fig:raise_time_analysis}. 
While a smoother pulse shape for the Rabi frequency slightly increases the time spent in the Rydberg state, such smoother pulses significantly reduce the bandwidth and the experimental requirements on fast switching of the laser pulses.

\section{Error-budgeting for Strontium-88
 setup}
\label{sec:error-budgeting}

In this section, we study the fidelity of the conditional phase gate taking into account the 
fundamental limitations of a Rydberg setup. We limit the analysis to the experimentally realistic 
Gaussian protocol (III.B) with a smooth switching on of the Rabi frequency. In the following, we take a rather conservative approach for the experimental parameters with room to 
improve the gate fidelity. The parameters 
are taken for a setup with strontium-88 atoms with the two-qubit states encoded in the two metastable
long-lived fine structure states $\ket{0} = \ket{5^{3}P_0}$ and $ \ket{1} = \ket{5^{3}P_{2}}$. Then, the qubit state $\ket{1}$ is coupled to the Rydberg state $\ket{r}=\ket{60^{3}S_1, m_J=1}$ through a single-photon process with Rabi frequency $\Omega_{0}$. Note that our choice to couple the state $\ket{1}$, which is higher in energy than $\ket{0}$, to the Rydberg state prevents ionization of the qubit. In the following, we provide the analysis for a Rabi frequency $\Omega_0 / 2 \pi = 10\;\text{MHz}$ and a wavelength $\lambda=323\;\text{nm}$. 
Note that a single-photon transition has the advantage of avoiding time-dependent light shifts and losses from an intermediate state. The van der Waals interaction between the atoms in the Rydberg state $\ket{r}$ can approximately be described by the coefficient $C_6/h=-154\; \text{GHz}\cdot\mu\text{m}^6$, as determined by the {\it pairinteraction} software~\cite{weber2017calculation}, using the quantum defects from Ref. \cite{robertson2021arc}.
For a realistic single-photon Rabi frequency $\Omega_0 / 2 \pi = 10\;\text{MHz}$ and interatomic distance of $R=3\;\mu\text{m}$, we obtain an interaction strength of $V/\hbar\Omega_0 =21.1$ as considered in Sec.~\ref{sec:different_shapes}.
For the calculation of the van der Waals coefficient, we assumed a quantization axes of the Rydberg state along the interatomic axis, such that a coupling to Rydberg states  $\ket{60^{3}S_1, m_J}$ with a different magnetic quantum number $m_{J}$ vanishes. Note that at these large distances, the interaction potential between the Rydberg states with different magnetic quantum numbers exhibits a very similar behavior~\cite{weber2017calculation}, such that a coupling is also strongly suppressed within moderate magnetic fields for different arrangements of the atoms.
Furthermore, all other Rydberg pair states that could be excited are separated by an energy gap larger 
than $1\;{\rm GHz}$, while the first level crossings only start to appear at distances $R\lesssim 2\;\mu\text{m}$.

Then, the smooth turning on of the Rabi frequency with a time scale $\kappa=5\; {\rm ns}$ guarantees 
that excitations into other Rydberg states are quenched and the latter can be ignored in the following. 
The lifetime of $\ket{r}$ is on the order of $1/\gamma= 50\;\mu\text{s}$ which is a conservative estimate, given that Ref.~\cite{madjarov2020} states $80\;\mu\text{s}$ for $n=61$. A precise determination of the lifetime would require multichannel quantum defect theory and is beyond the scope of this work \cite{vaillant2014multichannel}. The trapping of the atoms can be achieved at a triple magic wavelength, such that the two-qubit states, as well as the Rydberg state, have the same trapping potential \cite{meinert2021}. In the following, we use a trap frequency of $\omega_z/2\pi = 50\;\text{kHz}$ along the direction of the excitation laser and $\omega_x/2\pi = \omega_y/2\pi = 100\;\text{kHz}$ perpendicular to it. Then, there are three fundamental corrections to the idealized Hamiltonian in Eqs.~\eqref{eq:simple_hamiltonian} to \eqref{eq:hintsimple}, which will limit the gate fidelity: 
\begin{enumerate}
  \item[(i)] The Rydberg level $\ket{r}$ exhibits a finite lifetime $1/\gamma$.
  \item[(ii)] The recoil of the single-photon transition leads to a momentum transfer and an  energy shift.
  \item[(iii)] The interaction potential exhibits a position dependence and, therefore, the atoms experience a force in the Rydberg state.
\end{enumerate}
The last two points couple the qubit state $\ket{1}$ to the motional degree of freedoms and both effects depend on the motional cooling of the atoms
within the trapping potential of the optical tweezers. Therefore, we study these effects for different motional temperatures $T$.  

\begin{table}
  \sffamily
  \centering
  \begin{tabular*}{\linewidth}{l@{\extracolsep{\fill}}lrl}
    \toprule
    Max. Rabi frequency &$\Omega_{0}/2\pi$& 10 &MHz\\
    Trap frequency along $x,y$ &$\omega_x/2\pi = \omega_y/2\pi$ & 100 &kHz\\
    Trap frequency along $z$ &$\omega_z/2\pi$ & 50 &kHz\\
    Rydberg lifetime &$1/\gamma$ & 50 &$\mu$s\\
    Transition wavelength &$\lambda$ & 323 &nm\\
    VdW coefficient &$C_6/h$ & -154 &$\text{GHz}\cdot\mu\text{m}^6$\\
    Interatomic distance &$R$ & 3 &$\mu$m \\
    \botrule
  \end{tabular*}
    \caption{\emph{Experimental parameters for strontium-88.} The van der Waals (VdW) coefficient $C_6$, the lifetime of the Rydberg state $\ket{r}$ $1/\gamma$, and the wavelength $\lambda$ of the transition $\ket{1} \rightarrow \ket{r}$ are specific to our choice of the qubit and Rydberg state.}
    \label{tab:experimental_parameters}
\end{table}

In order to account for these phenomena, we have to modify the Hamiltonian describing the gate protocol. We recall our setup shown in 
Fig.~\ref{fig:setup} with the separation between the two atoms along the $x$-direction, while the driving laser is applied along the $z$-axis. 
Then, the coupling Hamiltonian $H_{0}$ to the Rydberg state is modified to 
\begin{equation}
    H_0= \hbar \sum_{i=1}^{2}   \left[ \frac{\Omega(t)}{2} \Big( \sigma_{i}^{+} e^{\mathrm{i} {\bf k} {\bf r}_{i}}+ \sigma_{i}^{-} e^{- \mathrm{i} {\bf k} {\bf r}_{i}} \Big) - \Delta(t) n_{i} \right]\;, 
\end{equation}
 where ${\bf k}=(0,0,2\pi / \lambda)^T$  with a wavelength $\lambda$ accounts for the momentum transfer of the single-photon transition to the Rydberg state,  
 and ${\bf r_i}=(x_{i}, y_{i}, z_{i})^T$  is the position  operator for the two atoms measured from the center of each trap. The trapping potential generated by the optical tweezers 
 at the triple magic wavelength is independent of the internal state and well described by a harmonic trap 
\begin{equation}
    H_{\rm \scriptscriptstyle trap}= \sum_{i=1}^{2} \left[ \frac{{\bf p}_{i}^{2}}{2 m} +  \frac{m \omega_{x}^2 }{2} x_{i}^2+ \frac{m \omega_{y}^2 }{2} y_{i}^2+ \frac{m \omega_{z}^2 }{2} z_{i}^2 \right] 
\end{equation}
with ${\bf p}_{i}$ the momentum operator of the atoms and  $m$  the mass of one strontium-88 atom. 
Note that the trapping frequencies in the optical tweezers  naturally satisfy $\omega_{x}=\omega_{y}> \omega_{z}$.
Furthermore, the interaction potential between the Rydberg states is well described by the van der Waals interaction
\begin{equation}
    H_{\rm \scriptscriptstyle int}= -\frac{C_6}{|{\bf R}+{\bf r}_{1}- {\bf r}_{2}|^{6}} n_{1}  n_{2} \, ,
\end{equation}
with ${\bf R}=(R, 0, 0)^T$ the separation between the two optical tweezers.  Finally, the spontaneous
emission from the Rydberg state is well accounted for within the framework of a Lindblad master
equation~\cite{breuer_petruccione} with decay rate $\gamma$. In general, the decay can happen into one 
of the two-qubit states or into an additional state outside of the computational basis. Here, we 
restrict the analysis to the latter case, which provides the lowest bound on the gate fidelity. Then, 
the decay from the Rydberg state can be accounted for by a non-hermitian contribution to the Hamiltonian with
\begin{equation}                                                                                              \label{eq:decoherence}
    H_{\rm \scriptscriptstyle decay}=  - \mathrm{i} \frac{\hbar \gamma}{2} \sum_{i=1}^{2}  n_{i} \, .
\end{equation}
The effective Hamiltonian $H= H_{0}+H_{\rm \scriptscriptstyle trap}+ H_{\rm \scriptscriptstyle int}+ H_{\rm \scriptscriptstyle decay}$ determines the full dynamics of the controlled-phase gate including these fundamental limitations. The full simulation of the Lindblad master equation via the density matrix or quantum trajectories is not required, because we can extract the decoherence effects via the norm reduced by Eq.~\eqref{eq:decoherence}.

\begin{figure}[t]
\centering
\includegraphics{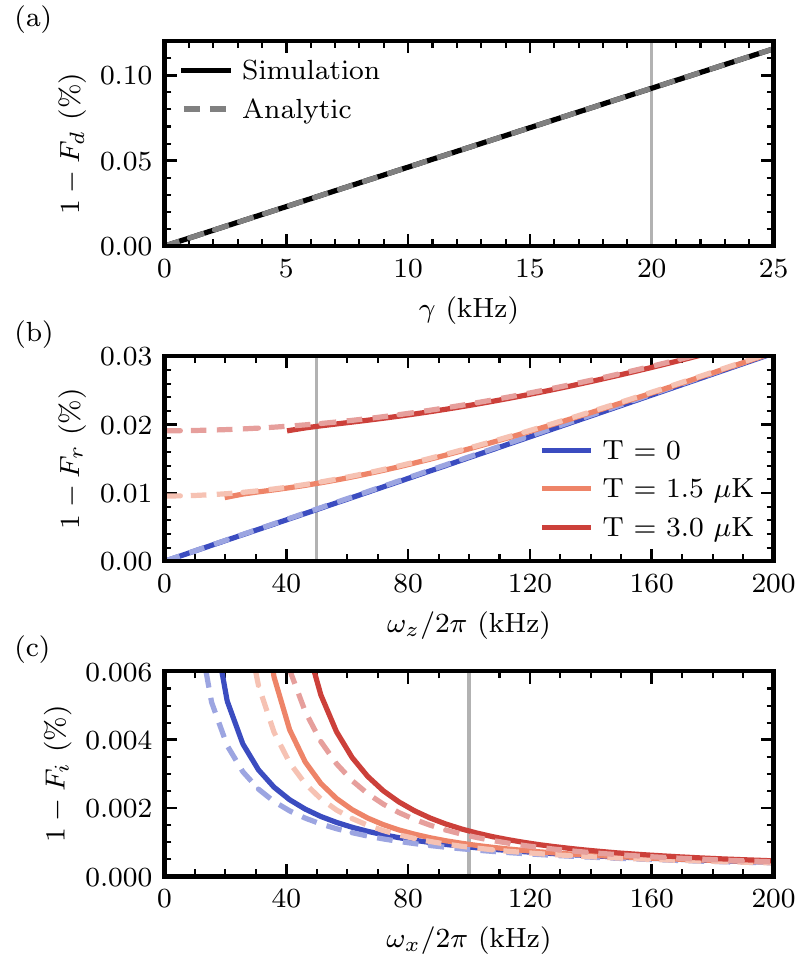}
\caption{\emph{Intrinsic infidelity.} For pulse (III.B), we analyze the infidelity caused by various effects that are intrinsic to realistic setups. The gray lines depict the parameters that we consider for our strontium-88 setup. (a) We simulate the Rydberg decay in a system where the atoms are at a fixed position. The dashed line depicts the analytic result of Eq.~\eqref{eqn:infidelity_decay}. The infidelity $1-F_d$ increases proportionally to the decay rate $\gamma$ and the average time spent in the Rydberg state $\overline{T}_{r}$. (b) The infidelity $1-F_r$ caused by photon recoil is studied in a system where the atoms are placed in one-dimensional harmonic traps along the direction of the laser $z$. The infidelity increases with the trap frequency $\omega_z$. We simulate the system for different temperatures $T$ of the atoms in the harmonic traps. The dashed curves show the analytic result of Eq.~\eqref{eqn:infidelity_recoil}. (c) The infidelity $1-F_i$, which is caused by the force due to van der Waals interaction, is simulated for a system where the atoms are placed in one-dimensional harmonic traps along $x$, i.e. the direction of the interatomic axis. The effect is only significant for trap frequencies $\omega_x \ll 100\;\text{kHz}$. The dashed curves are the analytic estimates from Eq.~\eqref{eqn:infidelity_interaction}.
}
\label{fig:intrinsic_effects}
\end{figure}

In the following, we analyze in detail the influence of each contribution to the infidelity $(1-F)$ for the quantum gate. Among the different measures to quantify gate performances, the fidelity  for the generation of the Bell state $\ket{B}=(\ket{00}+\ket{11})/\sqrt{2}$ is especially useful and widely used for Rydberg platforms \cite{gilchrist2005,theis2016,saffmann2021}. Another common measurement is the average gate fidelity \cite{nielsen2002}. However, it turns out that for our setup the Bell state 
fidelity  is more restrictive and provides a lower value than the average gate fidelity;  the latter has 
a higher weight in states, where the gate works perfectly.  In addition, the evaluation of the Bell state fidelity is numerically 
more efficient. Therefore, we use the Bell state fidelity to optimize our gates and analyze the different
error contributions.
We compare the value of the Bell state fidelity to the average gate fidelity for a few examples in Tab.~\ref{tab:infidelities}. 
Note that for the evaluation of the gate fidelity at zero motional temperature $T$, we start in a product state between the computational 
states and the motional ground state $\ket{\phi}$ of the harmonic traps, i.e., $\ket{\psi_{\rm \scriptscriptstyle inital}}= \ket{00} \otimes 
\ket{\phi}$. Then, we propagate the state, obtaining 
\begin{equation}
\ket{\psi_{\rm \scriptscriptstyle final}} = U_{1} U U_{2}\ket{\psi_{\rm \scriptscriptstyle initial}}\;,
\end{equation}
where $U$ denotes the time evolution under the Hamiltonian $H$, while
$U_{1}$ and $U_{2}$ are additional perfect single-qubit gates required for the generation of the Bell state $\ket{B}$ and correcting the phase factor $\varphi_{01}$ and $\varphi_{10}$. After tracing out the traps, the fidelity between the resulting reduced density matrix and the Bell state $\ket{B}$ is calculated \cite{saffmann2021}. For calculating the Bell state fidelity at non-zero motional temperatures, we follow the same procedure except for starting with the separable density matrix $\rho=\ket{00}\bra{00} \otimes \rho_\phi$, where $\rho_\phi$ is the density matrix of the thermal motional state.

First, we study the influence of spontaneous decay from the Rydberg state. We apply the pulse sequence (III.B) with a Gaussian shape and a smooth turning on of the Rabi frequency. 
In Fig.~\ref{fig:intrinsic_effects}(a), we plot the infidelity as a function of the decay rate.  We observe that the infidelity increases linearly with the decay rate and can be excellently approximated as
\begin{equation}
1-F_d = 3/4\;\overline{T}_r\gamma\;.
\label{eqn:infidelity_decay}
\end{equation}
This particular dependence on the time $\overline{T}_r$ introduced in Eq.~\eqref{eq:mean_time_rydberg_state} results from the fact that each of the four basis states $\ket{00}$,  $\ket{01}$, $\ket{10}$, and $\ket{11}$ enter with the same amplitude in the calculation of the Bell state fidelity.

Next, we extract the contributions on the infidelity caused by photon recoil, see Fig.~\ref{fig:intrinsic_effects}(b). The atoms are placed in one-dimensional harmonic traps along the $z$-direction, i.e., the direction of the laser beam. The infidelity increases with the trap frequency $\omega_z$.
This counter-intuitive behavior on the influence of the photon recoil in the harmonic oscillator potential, which has also been observed in \cite{saffmann2021}, can be well understood 
by a simple argument: starting in the ground state of the harmonic oscillator $\ket{1}\otimes \ket{\Omega}$, the photon recoil $e^{\mathrm{i} k z_{i}}$ acts as 
a displacement operator and generates the coherent state $\ket{\alpha}$ for the motional degree of freedom with  $\alpha= \mathrm{i} \sqrt{ \hbar/2 m \omega_z} k$. 
During the excitation of the atom in the Rydberg state, this state undergoes the coherent dynamics of the harmonic oscillator, and  after the time $\overline{T}_r$ is deexcited from the Rydberg level with the opposite displacement operator.
Therefore, the amplitude to
return to the ground state of the harmonic oscillator is determined by the overlap between  two coherent states
($\omega_z \overline{T}_{r} \ll 1$)
\begin{eqnarray}
\bra{\alpha} \ket{\alpha \mathrm{e}^{- \mathrm{i} \omega_z t}}   \approx  \exp\left[-  \mathrm{i} \frac{\hbar k^2}{2 m} \overline{T}_r - \frac{1}{2} \frac{\hbar k^2}{2 m} \omega_z \overline{T}_r^2 \right].
\label{eqn:infidelity_amplitude}
\end{eqnarray}
The first term accounts for an energy shift by the recoil energy and is taken 
into consideration as a slight change in the detuning for the optimal laser pulse, 
while the second term accounts for motional excitations and a reduced probability to return to the motional ground state. This analysis can be extended to a motional Fock state $\ket{n}$; note that the shift in detuning is independent on the excitation $n$. Such an analysis allows to determine the contribution to the infidelity from the photon recoil with a thermal density matrix at temperature $T$ for the initial motional state
\begin{equation}
 1-F_r=  \frac{15}{32} \frac{\hbar k^2}{2 m} \omega_z \overline{T}
 _r^2 \coth\left(\frac{\hbar \omega_z}{ 2 k_B T}\right).
 \label{eqn:infidelity_recoil}
\end{equation}
Here, the factor $15/32$ accounts for the influence of the coherent suppression in amplitude on the Bell state fidelity; in analogy to the factor $3/4$ in Eq.~(\ref{eqn:infidelity_decay}). Note, that the temperature dependence follows exactly the increase in momentum distribution for a thermal state. This observation allows for the interpretation, that the decrease in fidelity for increasing temperature follows from the additional Doppler broadening by the photon recoil. 
The estimation from Eq.~\eqref{eqn:infidelity_amplitude} agrees very well with our numerical optimized protocol where the shift in the detuning of $0.02\; {\rm MHz}$ originates from the first term in the exponential.
Note that this argumentation breaks down for very 
deep trapping potentials, where the motional sidebands can be resolved; the parameter regime of ion 
trap quantum computers \cite{bruzewicz2019trapped}. We simulate the system for 
different temperatures $T$ of the atoms in the harmonic traps, taking into account up to 10 harmonic oscillator modes. The results for small trap frequencies and high temperatures did not converge for the considered number of modes and are therefore not shown in the corresponding parameter regime. 
The temperature of $1.5\;\mu\text{K}$ corresponds to a ground state occupation of 96\% at trap frequency $\omega_x/2\pi=\omega_y/2\pi=100\;{\rm kHz}$ and 80\% for the weaker trap at $\omega_z/2\pi=50\;{\rm kHz}$. The full numerical simulations agree very will with the prediction by the above simplified analytical derivation, see Fig.~\ref{fig:intrinsic_effects}(b).

In Fig.~\ref{fig:intrinsic_effects}(c), we analyze the infidelity caused by the force due to the
van der Waals interaction. The atoms are placed in one-dimensional harmonic  traps along the 
$x$-direction, i.e., the direction of the interatomic axis. The influence of the van der Waals 
interaction increases for decreasing trap frequencies, but its contribution is further suppressed 
compared to the previous two effects at realistic trap frequencies $\omega_x /2 \pi \sim  100\,\text{kHz}$.
We can theoretically understand its contribution by the following analysis: during the gate, the state
$\ket{rr}$ is only  weakly occupied. Therefore, we replace in $H_{\rm \scriptscriptstyle int}$ 
the operator $n_1 n_2$ by its expectation value, and consider the lowest-order perturbation theory in $a_{x}/R$ with 
$a_x=\sqrt{\hbar/m \omega_x}$ being the harmonic oscillator length.  Then, the
harmonic oscillator states are perturbed by the time-dependent Hamiltonian
\begin{equation}
    H_{\rm \scriptscriptstyle pert} = - 6 V \: \frac{{\bf R}\cdot \left({\bf r}_{1}-{\bf r}_{2}\right)}{R^2}  \: T_{rr} \delta(t);
\end{equation}
note, that the quantum gate is performed on a time scale much faster than the characteristic time scale of the harmonic oscillator, and therefore the expecation value $\langle n_1 n_2 \rangle \approx T_{rr} \: \delta(t)$
is well described by a $\delta$-function with $T_{rr}$ the time spent in the state $\ket{rr}$ and $V=-C_6/R^6$. Using standard time dependent perturbation theory, we can derive the probability for transitions into different Fock states of the harmonic oscillators. Averaging over the thermal density matrix of the motional states, and accounting for its influence to on the gate fidelity (contributing a factor 3/16), we obtain
\begin{equation}
 1-F_i= \frac{27}{4} \left(\frac{T_{rr}   V}{\hbar}   \right)^2 \frac{ \hbar}{m \omega_x} \frac{1}{R^2} \coth\left(\frac{\hbar \omega_x}{ 2 k_B T}\right) \, ,
 \label{eqn:infidelity_interaction}
\end{equation}
The expression agrees very well with the numerical calculations, 
but at low trap frequencies, higher-order corrections become important, see Fig.~\ref{fig:intrinsic_effects}(c).

Therefore, the Rydberg decay represents the dominant source of error for the conservative 
experimental parameters of Tab.~\ref{tab:experimental_parameters}. The different
contributions sum up to a total Bell state infidelity of $0.101\%$ at zero-temperature,
which corresponds to a Bell state fidelity $F=99.899\%$,
see Tab.~\ref{tab:infidelities}. This value agrees excellently with a simulation 
of the full Hamiltonian, considering all contributions simultaneously, i.e., the different error sources seem to be independent of each other. For an increased 
temperature of $1.5\;\mu\text{K}$, the total infidelity becomes $0.105\%$ because of the increased contribution of the photon recoil.

\begin{table}
  \sffamily
  \centering
  \begin{tabular*}{\linewidth}{l@{\extracolsep{\fill}}cc@{\extracolsep{\fill}}cc}
    \toprule
    & \multicolumn{2}{l}{Bell state infidelity} & \multicolumn{2}{l}{Average gate infidelity} \\
    & 0 $\mu$K & 1.5 $\mu$K & 0 $\mu$K & 1.5 $\mu$K \\
    \colrule
    Rydberg decay  & 0.092\% & 0.092\% & 0.074\% & 0.074\% \\
    Photon recoil & 0.008\% & 0.011\% & 0.006\% & 0.009\%  \\
    VdW force & 0.001\% & 0.001\% & 0.001\% & 0.001\% \\
    \colrule
    Summed & 0.101\% & 0.105\% & 0.081\%  & 0.084\%   \\
    Full simulation & 0.101\% & - & 0.081\%  & - \\
    \botrule
  \end{tabular*}
    \caption{\emph{Error budgeting.} For our experimental parameters summarized in Tab.~\ref{tab:experimental_parameters}, the Rydberg decay is the dominant source of error. The photon recoil and the van der Waals (VdW) force are not the leading contributions. We checked for zero temperature that the sum of the different contributions to the infidelity is in excellent agreement with a simulation of the full Hamiltonian, considering all of the error sources simultaneously. The Bell state infidelity that we use throughout our work is typically larger than the average gate infidelity.}
    \label{tab:infidelities}
\end{table}

The analysis opens the path to determine the experimental challenges 
to further improve the gate fidelity: increasing the available Rabi frequency naturally 
reduces the time spent in the Rydberg state. In turn, it requires reducing the distance 
between the atoms in order to keep a strong interaction and Rydberg blockade. Furthermore, 
the switching time $\kappa$, in general, is already fixed by the available bandwidth to
manipulate laser pulses and is also limited by the condition to avoid transitions into 
higher Rydberg states. As an example, we illustrate the behavior on the gate fidelity by
increasing the  Rabi frequency to $\Omega_{0}/2\pi=40 {\rm MHz}$. It requires shortening the 
distance between the two atoms by a factor $4^{1/6}$ to $R= 2.38 \mu{\rm m}$ 
in order to keep  the same interaction strength. This decreases the infidelity from the 
decay $1-F_d$ and recoil $1-F_r$ by a factor of 4, but increases  the infidelity from 
the interaction $1-F_i$ by a factor $4^{1/3}\approx 1.58$. Furthermore, we keep the switching 
time for the Rabi frequency at $\kappa = 5 {\rm ns}$, i.e., $\kappa \Omega_{0}=1.24$. 
The latter increases the time spent in the Rydberg state by about 4\%, which affects
the contribution from the decay and the interaction. 
In combination, we can estimate the new infidelity for the process $1-F=0.027\%$ corresponding to a Bell state fidelity $F=99.973\%$ at $\Omega_{0}/2\pi=40 {\rm MHz}$, which again agrees well with the full numerical simulation.

\section{Summary and outlook}
\label{sec:conclusions}

In this work, we have investigated the implementation of a two-qubit controlled-phase (CZ) gate in a neutral atoms quantum platform. 
We have considered two neutral atoms trapped in optical tweezers at a fixed distance. The qubit states $\ket{0}$ and $\ket{1}$ are encoded into 
the electronic states of the atom. The logical state $\ket{1}$ of each qubit is coupled to a Rydberg state by a laser beam with a time-dependent 
Rabi frequency and a time-dependent detuning.

First, we have shown in an idealized description that the controlled-phase gate can be realized by adopting different detuning shapes. In particular, we have found a gate pulse sequences, which reduces the time spent in the Rydberg state by $10\%$  with respect to the protocol adopted in Ref.~\cite{levine2019}, and this reduction can be achieved for experimentally realistic pulses with smooth turning on of the Rabi frequency and smooth changes in the detuning. This reduction in time directly provides an improvement in the gate fidelity, as the leading 
mechanism to the infidelity is identified as losses from the Rydberg states. In addition, 
the smoothness of the pulses quenches accidental transitions into additional Rydberg states. 
We have adopted a Gaussian detuning to perform an error-budgeting analysis for a strontium-88 setup. For this analysis, we have considered a more sophisticated model for the system's  description. Indeed, we have taken into account intrinsic effects as the finite lifetime of the Rydberg state, as well as the trapping of the atoms in the optical tweezers,  the effect of photon recoil, and the microscopic interaction potential between the atoms. For rather conservative estimation of the parameters, we have demonstrated that a Bell state fidelity at zero-temperature of $F= 99.899\%$ (average gate fidelity $F=99.919\%$) at $\Omega_{0}/2\pi=10\;{\rm MHz}$ can be achieved, while an increase in the maximal Rabi frequency to $\Omega_{0}/2\pi= 40\;{\rm MHz}$ increases the Bell state fidelity to $99.973\%$.

{\it During the completion of our work, we became aware of related work demonstrating time-optimal gates with Rydberg atoms}~\cite{jandura2022}. {\it Note that the gate times for two-qubit gates are comparable to the protocols presented here. }


\acknowledgments We thank J\"urgen Stuhler (Toptica) for fruitful discussions. This project has received funding from the German Federal 
Ministry of Education and Research (BMBF) under the funding program quantum technologies - from basic research to market -- with the grant QRydDemo. S.M. acknowledges support by the Italian PRIN2017 and Fondazione CARIPARO.


%

\end{document}